\documentclass[journal]{IEEEtran}

\ifCLASSINFOpdf
\else
\fi

\makeatletter
\let\NAT@parse\undefined
\makeatother
\usepackage[colorlinks=true,linkcolor=black,citecolor=blue,urlcolor=blue,]{hyperref}
\usepackage[doipre={doi:~}]{uri}

\usepackage{amsmath}
\usepackage{latexsym}
\usepackage{algorithmic}
\usepackage{algorithm}

\usepackage{graphicx}
\ifCLASSOPTIONcompsoc 
\usepackage[caption=false,font=normalsize,labelfont=sf,textfont=sf]{subfig}
\else
\usepackage[caption=false,font=footnotesize]{subfig}
\usepackage{flushend}
\fi

\begin{document}
\title{Unsupervised Low-dose CT Reconstruction with One-way Conditional Normalizing Flows}

\author{Ran~An, Ke~Chen$^{*}$ and Hongwei~Li$^{*}$
\thanks{Ran An is with the School of Mathematical Sciences, Capital Normal University, Beijing 100048, CHINA, and also with the Centre for Mathematical Imaging Techniques, University of Liverpool, Liverpool L69 7ZL, UK.}
\thanks{Ke Chen is with the Department of Mathematics and Statistics, University of Strathclyde, Glasgow G1 1XQ, UK, and also with the Centre for Mathematical Imaging Techniques, University of Liverpool, Liverpool L69 7ZL, UK. (e-mail: k.chen@strath.ac.uk)}
\thanks{Hongwei Li is with the School of Mathematical Sciences, Capital Normal University, Beijing 100048, CHINA. (e-mail: hongwei.li91@cnu.edu.cn)}}

\markboth{An \MakeLowercase{\textit{et al.}}: Unsupervised Low-dose CT Reconstruction with One-way Conditional Normalizing Flows}%
{An \MakeLowercase{\textit{et al.}}: Unsupervised Low-dose CT Reconstruction with One-way Conditional Normalizing Flows}

\maketitle

\begin{abstract}
Deep-learning methods have shown promising performance for low-dose computed tomography (LDCT) reconstruction. However, supervised methods face the problem of lacking labeled data in clinical scenarios, and the CNN-based unsupervised denoising methods would cause excessive smoothing in the reconstructed image. Recently, the normalizing flows (NFs) based methods have shown advantages in producing detail-rich images and avoiding over-smoothing, however, there are still issues: (1) Although the alternating optimization in the data and latent space can well utilize the regularization and generation capabilities of NFs, the current two-way transformation strategy of noisy images and latent variables would cause detail loss and secondary artifacts; and (2) Training NFs on high-resolution CT images is hard due to huge computation. Though using conditional normalizing flows (CNFs) to learn conditional probability can reduce the computational burden, current methods require labeled data for conditionalization, and the unsupervised CNFs-based LDCT reconstruction remains a problem. To tackle these problems, we propose a novel CNFs-based unsupervised LDCT iterative reconstruction algorithm. It employs strict one-way transformation when performing alternating optimization in the dual spaces, thus effectively avoiding the problems of detail loss and secondary artifacts. By proposing a novel unsupervised conditionalization strategy, we train CNFs on high-resolution CT images, thus achieving fast and high-quality unsupervised reconstruction. Experiments on different datasets suggest that the performance of the proposed algorithm could surpass some state-of-the-art unsupervised and even supervised methods. 
\end{abstract}


\section{Introduction}
Computed tomography (CT) is a widely used medical imaging technique for detecting the inner structure of an object.  It's known that excessive X-rays might cause potential harm to the human body, such as cancer risk and genetic hazards \cite{lidestaahl2021estimated}. In medical diagnosis, the as low as reasonably achievable principle \cite{ALARA_solomon2020justification} was proposed to direct the use of X-ray doses. However, low-dose scanning will lead to noise in the projection data, thereby introducing severe noise and artifacts in the reconstructed image \cite{LDCT_goldman2007principles}. Therefore, LDCT reconstruction has always been a popular subject in the medical imaging community.

For LDCT, some denoising strategies on post-processing reconstructed images \cite{imgpost1_chen2009bayesian, imgpost2_kang2013image, imgpost3_green2016efficient} and pre-processing projection data \cite{sinopre1_zhang2019tensor, sinopre2_zhang2013bayesian,sinopre3_manduca2009projection} were proposed. Although easy and convenient, such methods ignore the consistency between image and projection data in the reconstruction procedure, usually causing blurring and secondary artifacts in the reconstructed image. Some other traditional methods introduce artificially designed priors (e.g. the total variation (TV) minimization \cite{TV_rudin1994total}) in iterative reconstruction \cite{SARTTV_yu2009compressed,epTV_tian2011low,OSCP_deng2019fast}, achieving better consistency and results. However, these methods often suffer from tedious tuning work for hyper-parameters and usually require a large number of iterations. In addition, the accuracy and scope of applicability of the artificial priors are also issues worthy of attention.

Compared with traditional methods, learning-based methods usually show advantages in reconstruction performance, which benefits from the priors learned in the ``Big Data''. Among them, the sparse coding method represented by K-SVD \cite{KSVD_elad2006image} learns the representation of image content and rebuilds the denoised image. Recently, deep-learning (DL) methods have shown powerful modeling and data-fitting abilities, and the DL-based LDCT reconstruction methods have also demonstrated success. For LDCT image post-processing, various neural network architectures such as FBPConvNet \cite{FBPConvNet_jin2017deep}, RED-CNN \cite{REDCNN_chen2017low}, DIRE \cite{DIRE_liu2019deep} and CTFormer \cite{wang2023ctformer} demonstrates good results. However, their purely post-processing strategy still ignores the image-data consistency and easily causes over-smoothing and detail loss in the reconstructed images. To keep a good consistency, some dual-domain denoising networks were proposed to process the projection data and image jointly in a whole framework, such as DRCNN \cite{drcnn_feng2021dual}, DDPNet \cite{DDPNet_ge2022ddpnet}, DuDoUFNet \cite{zhou2022dudoufnet} and DRONE \cite{wu2021drone}. Some other methods unroll regularized iterative algorithms into networks to perform adaptive training of the hyperparameters, achieving outstanding performance, representative methods include LEARN \cite{learn_chen2018learn} and PD-Net \cite{learnpd_adler2018learned}. Although these methods have shown exceptional performance in LDCT reconstruction, they follow a supervised learning framework and require a large amount of labeled data for training, which is difficult to fulfill in actual CT applications due to some ethical principles and the difficulties of achieving completely consistent repeat scans.

To deal with the reliance on labeled data, some unsupervised strategies were proposed for LDCT reconstruction. Some work simulate paired data with generative adversarial networks (GAN) \cite{GAN_creswell2018generative} training on unpaired data, representative methods including GAN-CIRCLE \cite{GAN-CIRCLE_you2019ct}, Cycle-Free CycleGAN \cite{Cycle-free_kwon2021cycle}, AdaIN-Based Tunable CycleGAN \cite{AdaIN_gu2021adain}, and IdentityGAN \cite{IdentityGAN_li2020investigation}. However, they mostly rely on the cyclic consistency loss, which usually causes unstable training of GAN. Moreover, ensuring the authenticity and accuracy of the generated paired data is not easy. On the other hand, some unsupervised LDCT image denoising methods based on the idea of  Noise2Noise (N2N) were proposed \cite{N2NT1_hasan2020hybrid,N2NT4_fang2021iterative,Half2Half_yuan2020half2half,S2MS_zhang2022s2ms}. However, these methods still require pairs of noisy images of the same scene, which are still difficult to collect in practical applications. Therefore, some self-supervised methods based on the noisy image itself were proposed, such as Noise2Sim \cite{N2Sim_niu2022noise} and Noise2inverse \cite{Noise2inverse_hendriksen2020noise2inverse}. Some others employ dual-domain denoising strategy to ensure consistency, such as ETSRP \cite{ETSRP_wagner2023benefit}, SSDDNet \cite{SSDD_niu2022self} and SDBDNet \cite{SDBD_an2024self}. These self-supervised methods show performance close to supervised ones. However, their pursuit of averaging makes them prone to cause over-smoothing and detail loss. To enhance the sharpness and details of the reconstructed image, GAN was utilized to process the output images of denoising networks, representative methods including DD-UNET \cite{ddunet_zheng2020dual} and CLEAR \cite{clear_zhang2021clear}. However, this approach still faces the collapse and accuracy issues of GAN. 

Generative models such as the normalizing flows (NFs) \cite{NFs_kobyzev2020normalizing} and diffusion models \cite{Diffusion_croitoru2023diffusion} have also been employed to learn priors in clean images to serve as regularization terms in iterative reconstructions. Wei \textit{et al.} proposed an alternating minimization algorithm with two-way transformation, i.e. to and from the latent variable by NFs to solve imaging inverse problems \cite{dunfs_wei2022deep}. Fabian \textit{et al.} proposed PatchNR \cite{PatchNR_altekruger2023patchnr} that trains NFs on normal-dose image patches and makes it the regularization term in iterative reconstruction. Similarly, He \textit{et al.} proposed EASEL \cite{EASEL_he2022iterative} with score-based diffusion model \cite{NSCNv2_song2020improved}, Liu \textit{et al.} and Xia \textit{et al.} proposed Dn-Dp \cite{DNDP_liu2023diffusion} and DPR-IR \cite{DPR-IR_xia2023diffusion} respectively based on denoising diffusion probabilistic model (DDPM) \cite{DDPM_ho2020denoising}, to conduct regularized iterative LDCT reconstruction. Such methods follow an excellent unsupervised strategy that only requires normal-dose images for training, which fits actual applications well. However, these methods still carry on some problems. Due to the huge computational burden, it is hard to train NFs on high-resolution images (such as those with a size of $512\times512$). Although some dimension-reduction \cite{Trumpets_kothari2021trumpets} and conditional probability learning \cite{CNFsLDCT_denker2020conditional} strategies have been proposed to solve this problem, the reduction of image dimension would inevitably lead to information loss, and existing conditionalization strategies often come with the requirement for labeled data, which makes the application of NFs in unsupervised LDCT reconstruction still faces great limitations. In addition, the two-way transformation of NFs between the noisy data and latent variables would lead to distribution bias thus causing detail loss and secondary artifacts in the reconstructed image. A well-recognized limitation of diffusion-model-based methods is that they often require quite a long reconstruction time as they usually need even more than a thousand iterative sampling steps. Although there exist fast sampling methods, e.g. DDIM \cite{song2020denoising} and DPM-solver \cite{lu2022dpm}, to reduce the iterative steps to 50 and even 10-20, their performance with inverse problems, especially for LDCT reconstruction, has yet to be verified.

This paper proposes a novel LDCT iterative reconstruction algorithm by improving the current NFs-based methods. Our method achieves high-quality and efficient reconstruction by employing an unsupervised framework that only requires normal-dose images for distribution learning. To better utilize the regularization and generation capabilities of NFs, we carry out dual-space alternating iterative reconstruction in the data and latent space. Instead of using the two-way transformation of NFs in the dual spaces, we propose a novel algorithm that only conducts strict one-way generation transformation thus effectively avoiding introducing secondary artifacts. Moreover, to efficiently train high-quality NFs on high-resolution images for unsupervised LDCT reconstruction, we propose a novel unsupervised conditionalization method and train conditional normalizing flows (CNFs), thus making our network easily act on high-resolution CT images. By utilizing the linearization technique and the ordered-subset simultaneous algebraic reconstruction technique (OS-SART) \cite{OSSART_wang2004ordered} for fast incremental reconstruction, our method achieves computation-efficient LDCT reconstruction. Our method performs similarly to supervised ones as an unsupervised framework and is faster than the popular diffusion-model-based iterative reconstruction methods. Experiments on two datasets demonstrate that our method effectively solves the main problems of the current NFs-based LDCT reconstruction methods. Compared to state-of-the-art unsupervised and even supervised LDCT reconstruction methods, our method shows promising performance. The main contributions of our work can be summarized as follows:
\begin{itemize}
    \item We propose a novel NFs-based unsupervised LDCT iterative reconstruction algorithm, which performs regularization in the data and latent spaces of the NFs and employs a strict one-way transformation. While giving full play to the regularization and generation capabilities of NFs, our method avoids the detail loss and secondary artifacts caused by the two-way transformation of noisy images.  
    \item We propose an unsupervised conditionalization strategy for the CNFs-based LDCT reconstruction problem that does not rely on paired training data. Based on this strategy we train unsupervised CNFs thus achieving efficient training on high-resolution CT images, and fast iterative reconstruction. To the best knowledge of us, this is the first time CNFs were incorporated into the LDCT reconstruction procedure.
    \item Experiments on different datasets demonstrate the high performance and fast reconstruction speed of our method compared with some state-of-the-art learning-based iterative reconstruction methods.
\end{itemize}

\section{Related Work}
\subsection{NFs-based LDCT Reconstruction}
The forward projection process of LDCT can be modeled in the following form:
\begin{equation}
    y = \mathbf{A}x + \eta,
    \label{eq_2.1}
\end{equation}
where $y\in R^{m}$ denotes the vectorized low-dose projection data, $x\in R^{n}$ is the vectorized ideal clean image, $\eta\in R^{m}$ is the low-dose noise introduced by low-dose scanning, and $\mathbf{A}\in R^{m \times n}$ signifies the known projection matrix.

Given the low-dose projection data $y$, the ideal reconstructed image $\hat{x}$ can be acquired by maximizing the following logarithmic probability:
\begin{equation}
    \hat{x} = \underset{x}\arg\max \log p(y|x) + \log p(x).
    \label{eq_2.2}
\end{equation}
where $p(y|x)$ is the posterior probability of $y$ given $x$, and $p(x)$ is the prior probability of $x$. Assuming $p(y|x)\sim \mathcal{N}\left(\mathbf{A}x, \sigma_0^2\right)$, equation (\ref{eq_2.2}) leads to a minimization problem:
\begin{equation}
    \hat{x} \in \underset{x}\arg\min \frac{1}{2\sigma_0^2}\|y-\mathbf{A}x\|_2^2-\log p(x).
    \label{eq_2.3}
\end{equation}
The prior probability $p(x)$ plays an important role in the above minimization problem, how to accurately and succinctly describe it is crucial. An emerging approach is to learn $p(x)$ within a large number of normal-dose images by the generative model NFs \cite{NFs_kobyzev2020normalizing}. In general, NFs learn a differentiable bijective mapping $\mathcal{F}_\theta=\mathcal{G}_\theta^{-1}$ parameterized by $\theta$ between the data distribution $p(x)$ and a simple distribution $p(z)$ such as the standard Gaussian distribution $p(z) \sim \mathcal{N}(0, 1)$. By a trained NFs $\mathcal{F}_\theta$, each image sample $x$ has a unique corresponding latent variable $z$ and can be bidirectionally mapped through the invertible network:
\begin{equation}
	\left\{
	\begin{array}{cr}
		\begin{aligned}
			z &= \mathcal{F}_\theta(x),\\ 
			x &= \mathcal{G}_\theta(z).
		\end{aligned}
	\end{array} 
	\right.
	\label{eq_2.4}
\end{equation}
Then, the prior probability $p(x)$ can be expressed as:
\begin{equation}
    \log p(x)=\log p(z)+\log \left|\operatorname{det} (\mathbf{D} \mathcal{F}_\theta(x))\right|,
    \label{eq_2.5}
\end{equation}
where $\mathbf{D} \mathcal{F}_\theta(x)$ denotes the Jacobian of $\mathcal{F}_\theta$ concerning $x$, and its determinant $\operatorname{det} (\mathbf{D}\mathcal{F}_\theta(x))$ accounts for the change in density from the transformation $\mathcal{F}_\theta$. 
It is noted that the second term on the right side of the equation (\ref{eq_2.5}) is a constant and $p(z) \sim \mathcal{N}(0, 1)$. Introducing (\ref{eq_2.5}) and $x = \mathcal{G}_\theta(z)$ into (\ref{eq_2.3}), we can get a new minimization problem about $z$:
\begin{equation}
    \hat{z}\in \underset{z}\arg\min \left\|y-\mathbf{A} \mathcal{G}_\theta(z)\right\|_2^2+\lambda\|z\|_2^2.
    \label{eq_2.6}
\end{equation}
where $\lambda$ is a parameter related to $\sigma_0^2$ that controls the regularization effect. 

\begin{figure}[!t]
	\centering
	\includegraphics[width=3.5in]{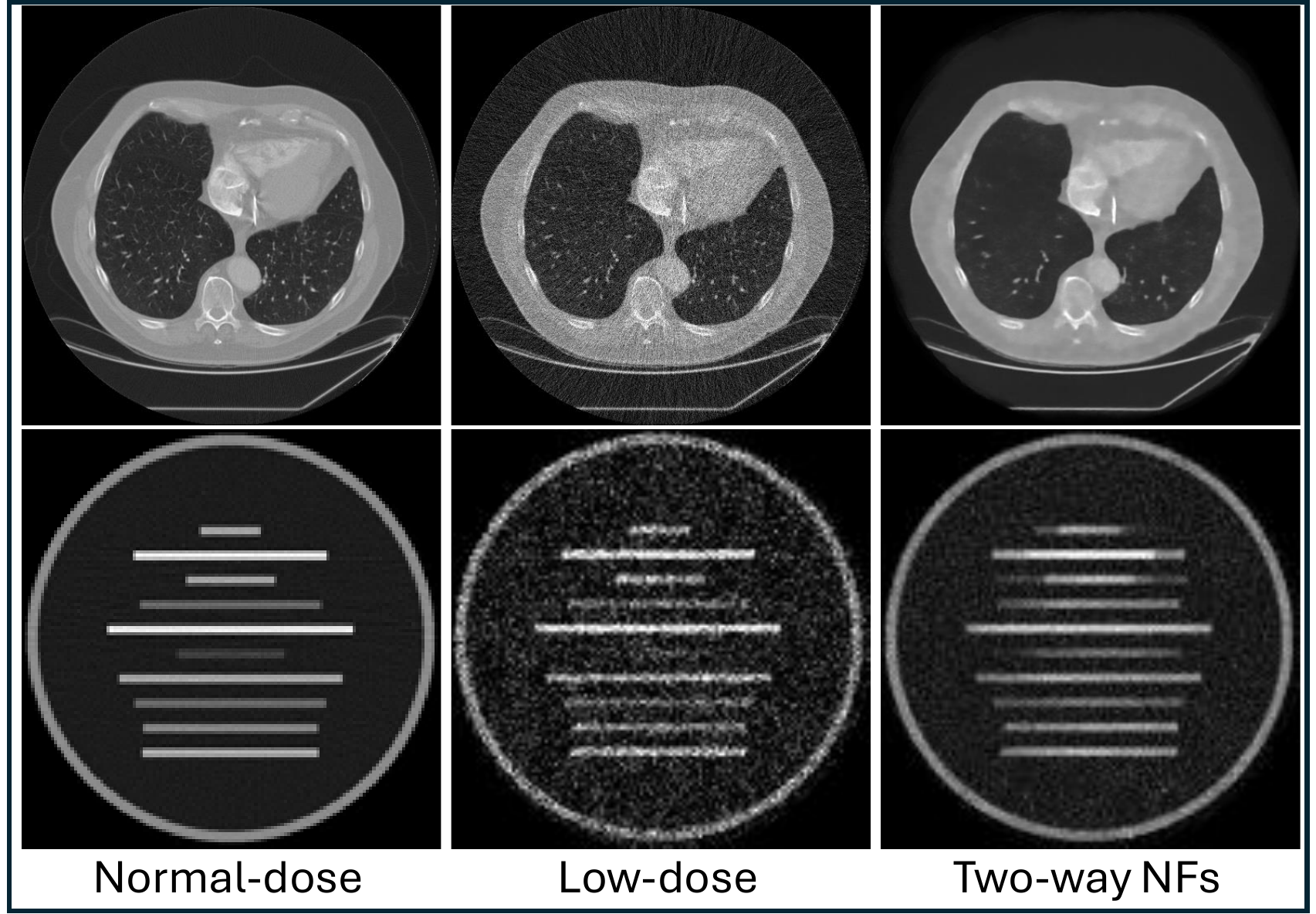}
    \vspace{-0.6cm}
	\caption{The reconstructed images of the two-way NFs. Compared to the normal-dose images, structure distortion and noise residuals can be easily observed.}
	\label{fig_1_TW-NFs}
\end{figure}

To solve the above problem (\ref{eq_2.6}), various methods have been proposed. In \cite{park2024solving} and \cite{helminger2021generic}, the authors propose to optimize $z$ directly in the latent space by gradient descent. Although this method well utilizes the generation ability of NFs and produces detail-rich images, it lacks constraints on the $x$-space and easily falls into a local optimum, usually leading to suboptimal reconstruction accuracy. In \cite{PatchNR_altekruger2023patchnr} and \cite{Trumpets_kothari2021trumpets}, the authors propose to perform optimization in the $x$-space. They make $\mathcal{G}_\theta(z)=x$ and $z=\mathcal{F}_\theta(x)$ in (\ref{eq_2.6}) and update $x$ with gradient descent too. This strategy can keep better reconstruction accuracy, however, it does not utilize the generation ability of NFs and easily causes excessive smoothing. To better use the regularization and generation capabilities of NFs, Wei \textit{et al.} \cite{dunfs_wei2022deep} proposed an alternating optimization strategy. They alternately update $x$ and $z$ in the data and latent space and conduct domain transfer by the two-way transformation of the NFs (i.e., $x=\mathcal{G}_\theta(z)$ and $z=\mathcal{F}_\theta(x)$). This method achieves effective dual-space alternating optimization and obtains promising results on the tasks involving natural images. However, it still presents drawbacks in LDCT reconstruction when high accuracy in the reconstructed image is crucial. Firstly, it does not constrain the consistency of the $x$ and $z$ in two adjacent iterations, which might adversely affect the data fidelity and thus the accuracy of the reconstructed image. More importantly, in the two-way transformation, the image to be transformed into the latent space is not a normal-dose image following the distribution of training samples but a noisy one. Therefore directly transforming it into the latent space by $z=\mathcal{F}_\theta(x)$ would cause a serious shift of the latent variable which shall affect subsequent iterations. In LDCT reconstruction, such problems would easily cause detail loss and introduce secondary artifacts in the reconstructed image, especially when the noise level is high. As shown in Figure \textbf{\ref{fig_1_TW-NFs}}, compared to the normal-dose images, the reconstructed images by the two-way NFs (TW-NFs) show severe artifacts and structure distortion. In \cite{dunfs_wei2022deep}, the authors propose to use the unfolding strategy to improve reconstruction performance, however, this additionally brings requirements for paired training data which is difficult to fulfill in clinical LDCT reconstruction.

\subsection{Conditional Normalizing Flows}
Although NFs exhibit good mathematical properties and enjoy efficient sampling and easy likelihood evaluation, their complex network structures with high-volume parameters make them quite compute-intensive. For generating small-size images such as those sized of $64\times 64$, NFs are more than capable. However, when trained on high-resolution images such as those sized of $512\times 512$, the memory scale and training time will become huge and unacceptable. As a case, it takes about a week to train a Glow \cite{Glow_kingma2018glow} (a classic NFs model) for 5-bits $256\times 256$ sized images with 40 GPUs \cite{Trumpets_kothari2021trumpets}. To achieve efficient training on high-resolution images, some new training strategies have been proposed to reduce the computation. Kothari \textit{et al.} \cite{Trumpets_kothari2021trumpets} propose injective flows Trumpets to reduce the dimension of the high-resolution images to a small size and train small-scale NFs. Although significantly reducing the computation cost and successfully training NFs on high-resolution images, the dimension reduction would inevitably result in information loss, leading to detail loss in the generated images. On the other hand, by introducing the already known information as conditions, conditional normalizing flows (CNFs) \cite{CNFs_winkler2019learning} were proposed to learn conditional probability instead of the complete probability $p(x)$. CNFs introduce conditions into the training phase thus helping to ease the training on high-resolution images. Furthermore, CNFs can also introduce conditions into inference and provide information for image generation, making it more efficient and meeting the specific requirements of practical applications. Similar to (\ref{eq_2.4}), the bidirectional process of CNFs with the condition $c$ can be expressed as:
\begin{equation}
\left\{
\begin{array}{cr}
\begin{aligned}
    z &= \mathcal{F}_\theta(x,c),\\ 
    x &= \mathcal{G}_\theta(z,c).
\end{aligned}
\end{array} 
\right.
\label{eq_2.12}
\end{equation}

Although CNFs can achieve training on high-resolution images and improve generation efficiency, the application of CNFs still presents challenges in unsupervised LDCT reconstruction. Current CNFs-based LDCT reconstruction methods often rely on paired data for their conditionalization \cite{CNFsLDCT_denker2020conditional, CTFlow_wei2023ctflow}, which is hard to implement in an unsupervised framework. Besides, they directly set the condition as the low-dose image, which might introduce misleading information such as noise and artifacts into the inference procedure. Although Wolf \textit{et al.} \cite{wolf2021deflow} proposed a down-sampling conditionalization method for unpaired image denoising, this strategy cannot work well when the low-dose noise and artifacts are severe. On the other hand, existing methods usually utilize CNFs for pure image generation rather than in a reconstruction procedure \cite{CNFsLDCT_denker2020conditional, CTFlow_wei2023ctflow, liu2022learning}, which would cause inconsistency between the projection data and reconstructed image, leading to bad accuracy of the image structures.

\section{Methods}
Based on CNFs, we propose an end-to-end unsupervised iterative LDCT reconstruction algorithm. Our method consists of three key ingredients: the unsupervised conditionalization strategy, the one-way iterative LDCT reconstruction algorithm, and the network of the CNFs. In this section, we will introduce these three parts in detail.

\subsection{Unsupervised Conditionalization Strategy}
In the unsupervised training phase of CNFs, only normal-dose images are accessible, therefore the conditions should come from the normal-dose images themselves to make it an unsupervised framework. On the other hand, in the inference stage, the only known information will be the low-dose projection data, which also means the condition should come from the low-dose data itself. To fit the training setup, the conditions in the two stages should be consistent (i.e. following the same distribution). Therefore, we propose the following two key points that the conditions should meet:
\begin{itemize}
    \item The conditions should contain most of the structure and feature information in the ideal clean image, while not carrying redundant information such as excessive noise and artifacts.
    \item The two conditions corresponding to the normal-dose and low-dose data of the same object should have a high degree of similarity and, ideally should be the same.
\end{itemize}

Based on the above two key points, we propose a new conditionalization method for CNFs-based unsupervised LDCT reconstruction. Specifically, for the low-dose projection data $y$, to get the intuitive image information, we first reconstruct it into an image $x$ with a reconstruction operator $\mathcal{R}$:
\begin{equation}
    x = \mathcal{R}(y).
    \label{eq_3.8}
\end{equation}
Then we remove most of the noise and artifacts in the image by a plug-and-play denoiser $\mathcal{D}$ (e.g. BM3D \cite{BM3D_dabov2007image}, NLM \cite{NLM_buades2011non} and DnCNN \cite{DnCNN_zhang2017beyond}) and a high-frequency-filter wavelet reconstruction operator $\mathcal{W}$:
\begin{equation}
    c^{'} = \mathcal{W}(\mathcal{D}(x)).
    \label{eq_3.9}
\end{equation}
Finally we add low-level Gaussian noise $n\sim \mathcal{N}\left(0, \sigma_1^2\right)$ to $c^{'}$ to enhance the robustness of the condition:
\begin{equation}
    c = c^{'} + n,
    \label{eq_3.10}
\end{equation}
where $\sigma_1$ is the standard deviation of the added noise. The whole process $\mathcal{C}$ for generating the condition $c$ from the projected data $y$ can be expressed as:
\begin{equation}
    c = \mathcal{C}(y) = \mathcal{W}(\mathcal{D}(\mathcal{R}(y))) + n.
    \label{eq_3.11}
\end{equation}
In the training stage, we can substitute $\mathcal{R}(y)$ in the above equation with the normal-dose image $x$ and use the same $\mathcal{W}$ and $\mathcal{D}$ to obtain the conditions of the normal-dose images. It is worth mentioning that although we use the denoiser on the normal-dose image here, it will not seriously destroy the image structures and details, and can produce a condition consistent with the low-dose image.

\begin{figure}[!t]
	\centering
	\includegraphics[width=3.5in]{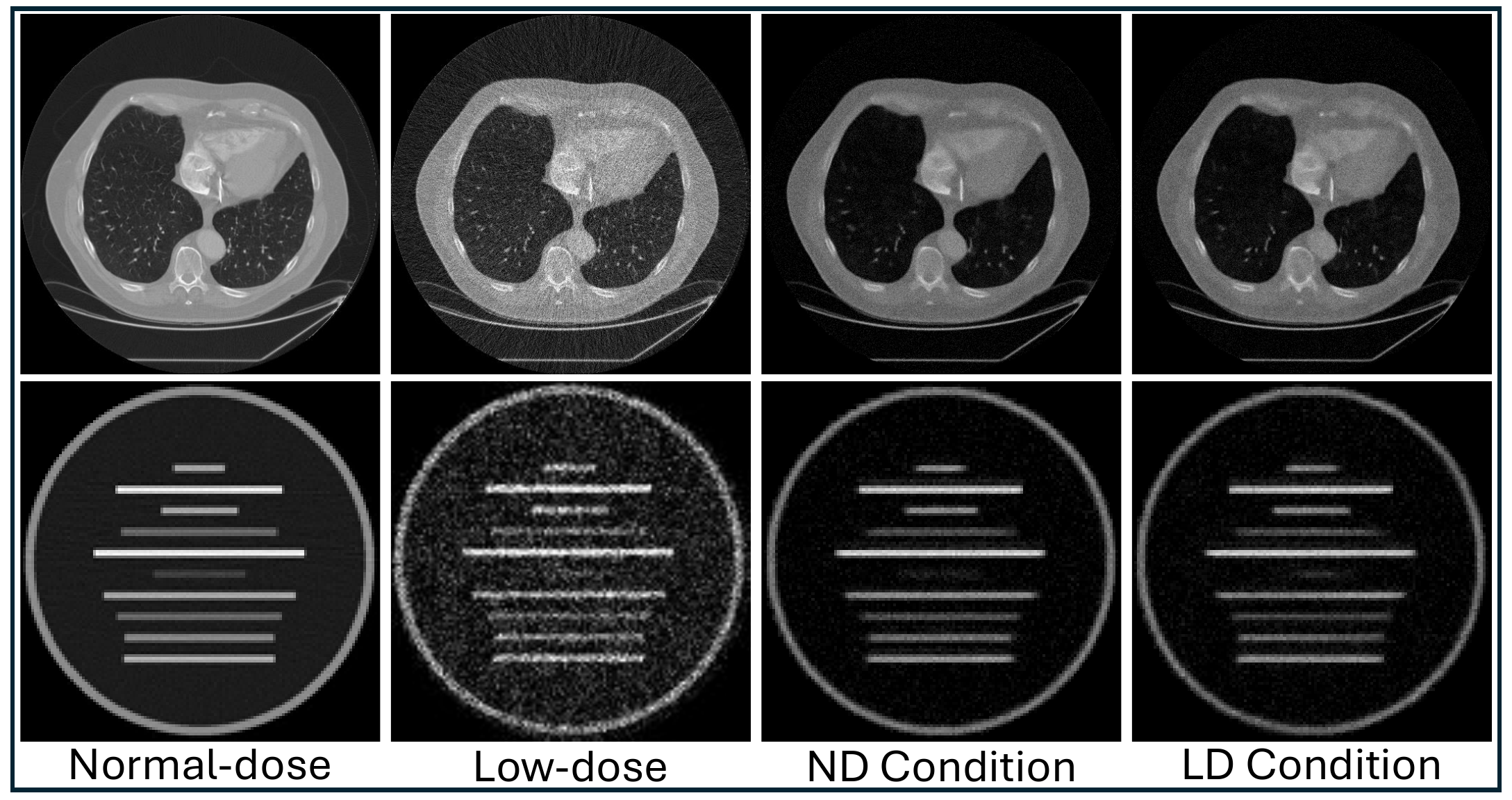}
    \vspace{-0.6cm}
	\caption{Condition examples generated with the BM3D denoiser.}
	\label{fig_2_Conditions}
\end{figure}

By our conditionalization method, we can get the conditions of both the normal-dose and low-dose data for unsupervised LDCT reconstruction. To evaluate the quality of the obtained conditions and adjust the parameters (e.g. The noise level parameter of BM3D or NLM) in our conditionalization method, a pair of normal-dose and low-dose data is needed, the criteria are set as the SSIM \cite{PSNR_hore2010image} between the paired conditions and the SSIM between them and the clean normal-dose image. Generally, a group of high SSIM means good quality of the conditions. As examples, conditions generated by the BM3D denoiser are shown in Figure \textbf{\ref{fig_2_Conditions}}.

\subsection{One-way CNFs LDCT Reconstruction Algorithm}
To tackle the problems of the current dual-space alternating optimization methods, we propose a novel iterative reconstruction algorithm. Our method has two main improvements: (1). To keep the consistency of $x$ and $z$ in iterations, we introduce constraint terms and update them separately; (2). Instead of the two-way transformation strategy, we adopt a strict one-way transformation between the data $x$ and the latent variable $z$ in the alternating optimization, thus effectively avoiding secondary artifacts. In detail, by introducing the trained CNFs $\mathcal{G}_\theta$, we can reformulate the problem (\ref{eq_2.6}) into the following double-variable minimization problem with a constraint term:
\begin{equation}
    \begin{gathered}
        (\hat{x}, \hat{z})=\underset{x,z}\arg\min \left\|y-\mathbf{A}x\right\|_2^2 + \lambda\|z\|_2^2 + \sigma\left\|x-\mathcal{G}_\theta(z,c)\right\|_2^2,
    \end{gathered}
    \label{eq_3.12}
\end{equation}
where $\sigma$ is a manual parameter to constrain the proximity between $x$ and $\mathcal{G}_\theta(z,c)$. With the incremental proximal-point method, this problem can be separated into the following two sub-problems:
\begin{equation}
    \begin{aligned}
    x^{n+1}=\underset{x}\arg\min \ \left\|y-\mathbf{A}x\right\|_2^2 &+ \sigma\left\|x-\mathcal{G}_\theta(z^{n},c)\right\|_2^2 \\
    &+ r_{1}\left\|x-x^{n}\right\|_2^2,
    \label{eq_3.13}
    \end{aligned}
\end{equation}
and
\begin{equation}
    \begin{aligned}
    z^{n+1}=\underset{z}\arg\min \ \lambda\|z\|_2^2 &+ \sigma\left\|x^{n+1}-\mathcal{G}_\theta(z,c)\right\|_2^2 \\
    &+ r_{2}\left\|z-z^{n}\right\|_2^2,
    \label{eq_3.14}
    \end{aligned}
\end{equation}
where $r_{1}$ and $r_{2}$ are two manual parameters which control the proximity of $x$ and $z$ in two adjacent iterations. It is worth noting that we only introduce the generation process $x=\mathcal{G}_\theta(z,c)$ of the CNFs here and conduct a strict one-way transformation from the latent variable $z$ to the image $x$, so that the problems of detail loss and secondary artifacts come with the two-way transformation are neatly avoided.

The sub-problem (\ref{eq_3.13}) admits a closed-form solution:
\begin{equation}
    x^{n+1} = (\mathbf{A}^{T}\mathbf{A} + \sigma\mathbf{I} + r_{1}\mathbf{I})^{-1}(\mathbf{A}^{T}y + \sigma \mathcal{G}_\theta(z^{n},c) + r_{1}x^{n}).
    \label{eq_3.15}
\end{equation}
To solve the sub-problem (\ref{eq_3.14}), we need the derivative of $\mathcal{G}_\theta(z,c)$ with respect to $z$. However, even though the CNFs are differentiable, it is hard to express the derivative of such a network to a variable. To deal with this problem, we linearize the second term in (\ref{eq_3.14}) which is hard to differentiate. Let $s(z) = \left\|x^{n+1}-\mathcal{G}_\theta(z,c)\right\|_2^2$, based on the diffeomorphism property of the CNFs $\mathcal{G}_\theta$, $s(z)$ can be approximated by its first-order Taylor expansion at $z^{n}$:
\begin{equation}
    \begin{aligned}
        s(z)\approx \ &s(z^{n}) + <s^{'}(z^{n}), (z-z^{n})> =\left\|x^{n+1}-\mathcal{G}_\theta(z^{n},c)\right\|_2^2\\ 
        &- 2\mathcal{G}^{'}_\theta(z^{n},c)(x^{n+1}-\mathcal{G}_\theta(z^{n},c))(z-z^{n}).
    \end{aligned}
    \label{eq_3.16}
\end{equation}
By substituting this approximation into the sub-problem (\ref{eq_3.14}), the solution $z^{n+1}$ is given by:
\begin{equation}
    z^{n+1} = \dfrac{\sigma \mathcal{G}^{'}_\theta(z^{n},c)(x^{n+1}-\mathcal{G}_\theta(z^{n},c)) + r_{2}z^{n}}{\lambda + r_{2}}, 
    \label{eq_3.17}
\end{equation}
where the $\mathcal{G}^{'}_\theta(z^{n},c)(x^{n+1}-\mathcal{G}_\theta(z^{n},c))$ can be accessed by the automatic derivation of the following loss to $z^{n}$ with the \textbf{Pytorch} functions: 
\begin{equation}
    \begin{gathered}
        \mathcal{L} = -\frac{1}{2}\left\|x^{n+1}-\mathcal{G}_\theta(z^{n},c)\right\|_2^2.
    \end{gathered}
    \label{eq_3.18}
\end{equation}

In summary, the iteration process of our unsupervised one-way CNFs reconstruction algorithm can be expressed as:
\begin{equation}
\left\{
\begin{array}{cr}
\begin{aligned}
    x^{n+1} &= (\mathbf{A}^{T}\mathbf{A} + \sigma\mathbf{I} + r_{1}\mathbf{I})^{-1}(\mathbf{A}^{T}y + \sigma \mathcal{G}_\theta(z^{n},c) + r_{1}x^{n}), \\ 
    z^{n+1} &= \dfrac{\sigma \mathcal{G}^{'}_\theta(z^{n},c)(x^{n+1}-\mathcal{G}_\theta(z^{n},c)) + r_{2}z^{n}}{\lambda + r_{2}}, \\ 
    \hat{x} &= \mathcal{G}_\theta(\hat{z},c) = \mathcal{G}_\theta(z^{\mathrm{K}},c),
\end{aligned}
\end{array} 
\right.
\label{eq_3.19}
\end{equation}
where $\hat{z}$ is the final output $z$ of the $\mathrm{K}_{th}$ iteration, $\hat{x}$ is the final reconstructed image generated by the CNFs with $\hat{z}$.

\begin{figure*}[!t]
	\centering
	\includegraphics[width=7.0in]{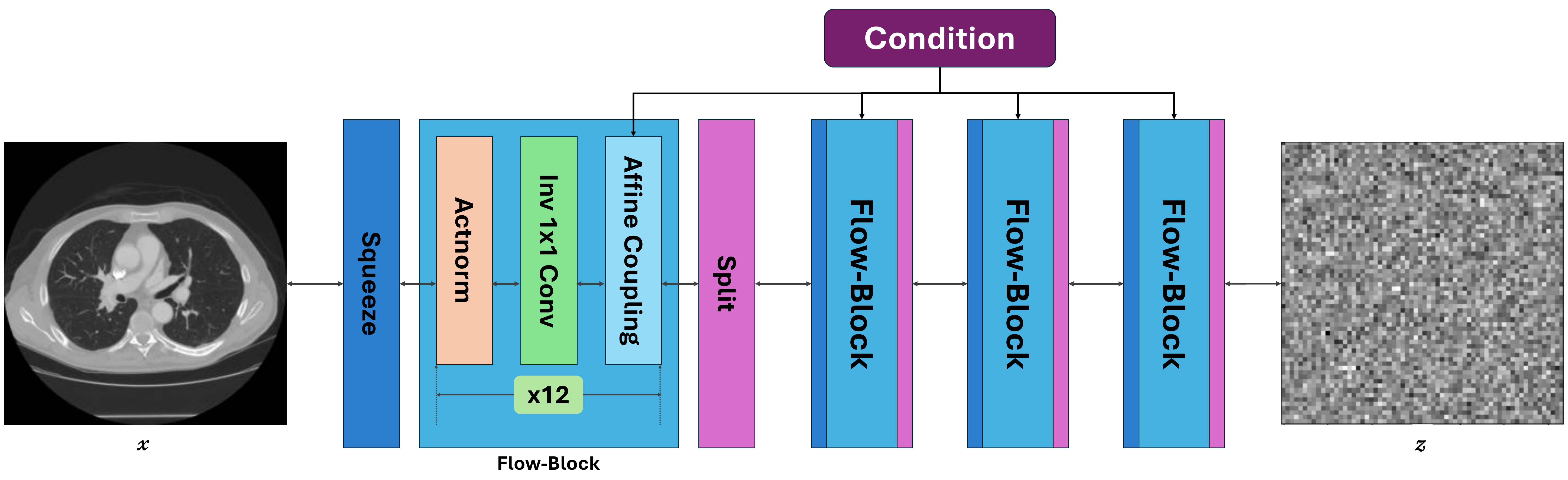}
    \vspace{-0.4cm}
	\caption{The network structure of our conditional normalizing flows.}
	\label{fig_3_cnfs}
\end{figure*}

In the iteration of $x^{n+1}$, the inverse of $(\mathbf{A}^{T}\mathbf{A} + \sigma\mathbf{I} + r_{1}\mathbf{I})$ is required for the updating. However, since the projection matrices $\mathbf{A}$ corresponding to each projection angle are different, using the matrix-form $\mathbf{A}$ in computation is impractical due to the huge storage and time consumption. On the other hand, if we use the operator form of $\mathbf{A}$, solving the inverse of $(\mathbf{A}^{T}\mathbf{A} + \sigma\mathbf{I} + r_{1}\mathbf{I})$ will be difficult. Therefore, we employ a two-step incremental reconstruction strategy to calculate an approximation to  $x^{n+1}$. In detail, we first reconstruct an intermediate image $x^{n+\frac{1}{2}}$ by the ordered-subset simultaneous algebraic reconstruction technique (OS-SART) \cite{OSSART_wang2004ordered}  with $x^{n}$ as the initial value:
\begin{equation}
    x^{n+\frac{1}{2}} = \textbf{OS-SART}(x^{n}, y, \omega)
    \label{eq_3.20}
\end{equation}
where $\omega$ is the relaxation factor. This step can be seen as minimizing the first term in (\ref{eq_3.13}). It is worth mentioning that using OS-SART to solve the fidelity term here is a more reasonable way than minimizing the $l_2$ norm, because modeling the low-dose noise of the projection data with a fix-variance Gaussian distribution is not accurate, while the reconstruction process of OS-SART can respect the noise characteristics better. In the second step, the solution $x^{n+1}$ can be expressed as the following form without any matrix inversion:
\begin{equation}
    x^{n+1} = \dfrac{x^{n+\frac{1}{2}} + \sigma \mathcal{G}_\theta(z^{n},c) + r_{1}x^{n}}{1 + \sigma + r_{1}}.
    \label{eq_3.21}
\end{equation}
This step can be seen as minimizing the remaining terms in (\ref{eq_3.13}). For the OS-SART, we set the number of iterations as $1$ and use a small and carefully tuned $\omega$, aiming for a good balance between the data fidelity and CNFs priors which could avoid structure distortion and inadequate denoising. Overall, the flow of our one-way CNFs LDCT reconstruction algorithm can be summarized as Algorithm \textbf{\ref{algo_1}}.

\begin{algorithm}[t]
	\caption{One-way conditional normalizing flows \\ (OW-CNFs) unsupervised LDCT reconstruction algorithm}
        \label{algo_1}
	\begin{algorithmic}[1]
		\STATE \textbf{Input:} The noisy projection data $y$, the hyperparameters $\lambda$, $\sigma$, $r_{1}$ and $r_{2}$, the relaxation parameter $\omega$ of OS-SART, the random Gaussian initialization $z^{0}$ and $x^{0}=\mathcal{G}_\theta(z^{0},c)$, and the number of iterations $\mathrm{K}$.
		\STATE \textbf{Operators:} The NFs $\mathcal{G}_\theta$ and the \textbf{OS-SART}.
        \STATE \textbf{Output:} The final reconstructed image $\hat{x}$.
		\STATE \quad \textbf{for} $n = 0, ..., \mathrm{K}-1$:
		\STATE \quad \quad $x^{n+\frac{1}{2}} = \textbf{OS-SART}(x^{n}, y, \omega)$
		\STATE \quad \quad $x^{n+1} = \dfrac{x^{n+\frac{1}{2}} + \sigma \mathcal{G}_\theta(z^{n},c) + r_{1}x^{n}}{1 + \sigma + r_{1}}$
		\STATE \quad \quad $z^{n+1} = \dfrac{\sigma \mathcal{G}^{'}_\theta(z^{n},c)(x^{n+1}-\mathcal{G}_\theta(z^{n},c)) + r_{2}z^{n}}{\lambda + r_{2}}$.
        \STATE \quad \textbf{end}
        \STATE \quad $\hat{x} = \mathcal{G}_\theta(z^{\mathrm{K}})$
		\STATE \textbf{return} $\hat{x}$
	\end{algorithmic}
\end{algorithm}

\subsection{Network}
The overall network of our CNFs is composed of a backbone NFs network and a condition module. The NFs consists of $4$ flow blocks each containing $12$ tandem combinations of the Actnorm layer, the invertible $1\times1$ convolutional layer and the affine coupling layer, similar to the famous normalizing flows Glow \cite{Glow_kingma2018glow}, and the number of the feature channels in each convolutional layer is $512$. In the condition module, the condition $c$ will be incorporated into each affine coupling layer of the backbone NFs by concatenation with the mainstream features. The overall network structure of our CNFs is shown in Figure \textbf{\ref{fig_3_cnfs}}.

\section{Experiments} 

\subsection{Experimental Setup}
To evaluate the performance of the proposed one-way conditional normalizing flows (\textbf{OW-CNFs}) unsupervised LDCT reconstruction algorithm, we performed experiments on two datasets. We evaluated the reconstructed images of our method by visual effects and quantitative indicators, and compared the results with several state-of-the-art deep-learning methods, including some popular unsupervised generative-model-based methods such as PatchNR \cite{PatchNR_altekruger2023patchnr}, EASEL \cite{EASEL_he2022iterative} and DPR-IR \cite{DPR-IR_xia2023diffusion}, and supervised LDCT denoising networks such as RED-CNN \cite{REDCNN_chen2017low} and CTFormer \cite{wang2023ctformer}.

\begin{figure}[!t]
	\centering
	\includegraphics[width=3.5in]{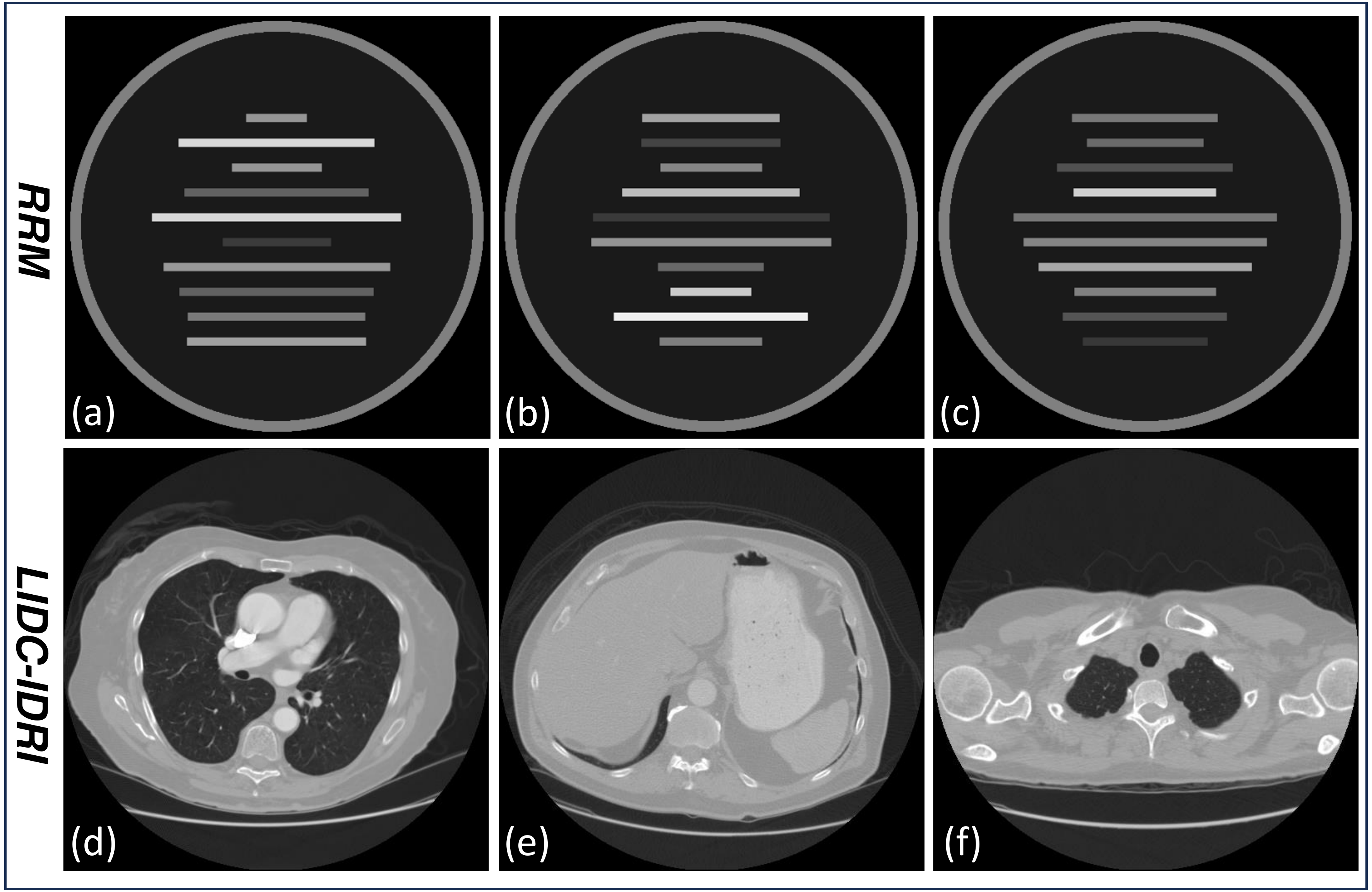}
    \vspace{-0.6cm}
	\caption{Some examples of the \textit{"RRM"} dataset ((a)-(c)) and \textit{"LIDC-IDRI"} dataset ((d)-(f)).}
	\label{fig_4_exmples}
\end{figure}

\begin{figure*}[!t]
	\centering
	\includegraphics[width=7.2in]{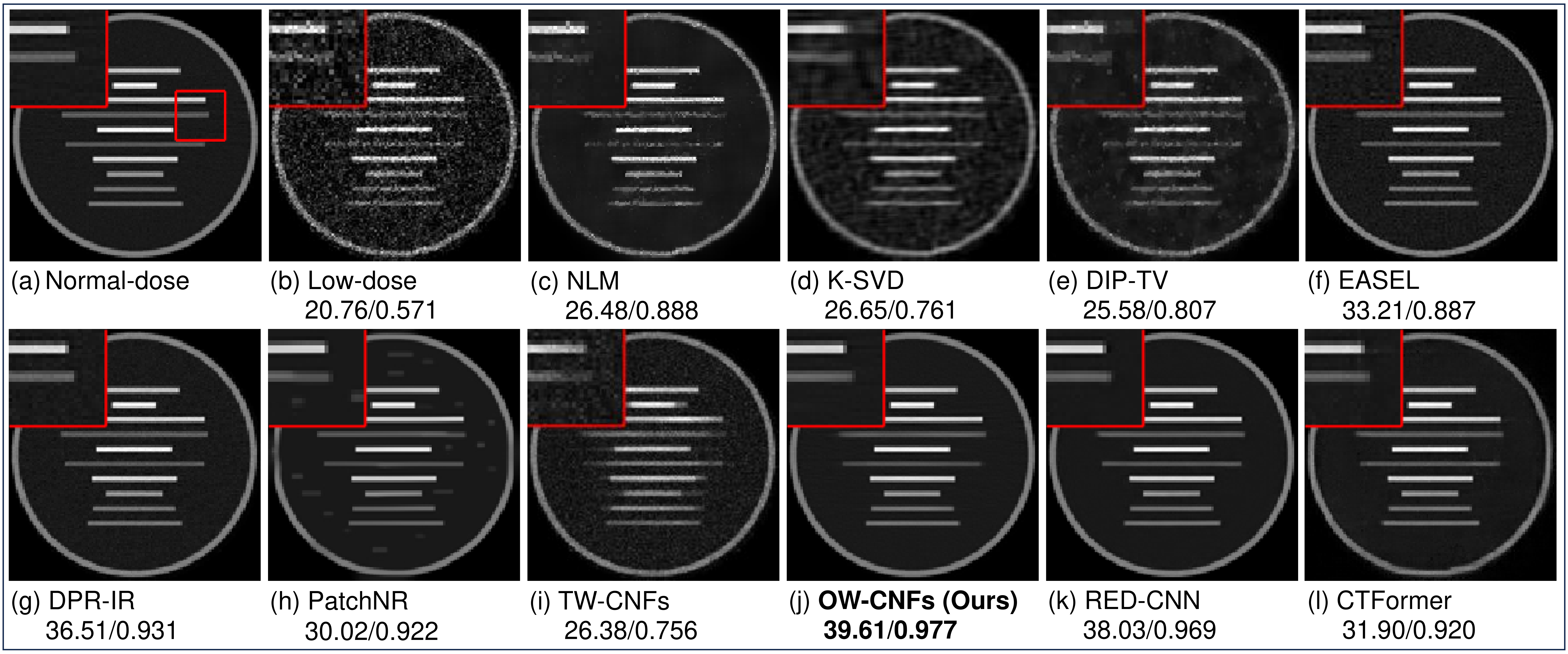}
    \vspace{-0.6cm}
	\caption{Reconstruction results of each method on the \textit{"RRM"} dataset at dose $I_{0}=1\times{10^3}$. The display window of the gray value range is set to $[0,1.0]$.}
	\label{fig_5_RRM}
\end{figure*}

The experiments were conducted on two datasets: (1) The random rectangle models (RRM) dataset and (2) The LIDC-IDRI dataset \cite{idri_armato2011lung}. The RRM dataset is home-made, consisting of a randomly generated series of images, each containing a large ring and ten parallel stripe rectangles with different lengths and gray values, and some examples are shown in Figure \textbf{\ref{fig_4_exmples}}(a)-(c). The reconstruction performance of different methods can be directly recognized from the reconstruction of those rectangular stripes. This dataset contains $1024$ images for training, $128$ images for validation and $32$ for testing, every image has a size of $128\time128$ and a gray value range of $[0, 1.0]$. The LIDC-IDRI dataset is a public CT image dataset comprising $241689$ normal-dose slices of $1012$ patients. We randomly picked $4409$ images from this dataset as the training set, $272$ for validation, and $30$ for testing. Some examples are shown in Figure \textbf{\ref{fig_4_exmples}}(d)-(f). All the picked CT images have a Hounsfield units (HU) range of $[-1024,2048]$.

The normal-dose and low-dose projection data were generated based on the above datasets with a simulated projection algorithm under a fan-beam imaging system. For the RRM dataset, we use 360 projection views uniformly distributed in the range $[0, 2\pi]$ and a linear detector with 256 cells. For the LIDC-IDRI dataset, we set both the projection views and the number of detector cells as 1000. We generated the clean projection data and added Poisson noise to the incident rays to simulate the low-dose projection data:
\begin{equation}
    y_{n} = -\ln(\frac{I_{d}}{I_{0}}), I_{d} \sim \mathrm{Poisson}\{I_{0}\times e^{-y_{c}}\},
	\label{eq_4.22}
\end{equation}
where $y_{c}$ is the clean projection data and $y_{n}$ is the noisy low-dose one. $I_{0}$ and $I_{d}$ are the number of incident and collected photons, respectively. Generally, a smaller $I_{0}$ means a lower dose and the projection data shall be more noisy. Based on this principle, we set $I_{0}=1\times{10^6}$ to simulate the normal-dose projection data, and the normal-dose images for reference were reconstructed on these projection data by OS-SART. For generating the low-dose projection data, we set $I_{0}=1\times{10^3}$ on the RRM dataset, and $I_{0}=1 \times {10^4}$ on the LIDC-IDRI dataset. The projection operator and OS-SART were coded with CUDA kernels wrapped by the \textbf{Cupy} library (\url{https://github.com/cupy/cupy}).

\begin{figure*}[!t]
	\centering
	\includegraphics[width=7.2in]{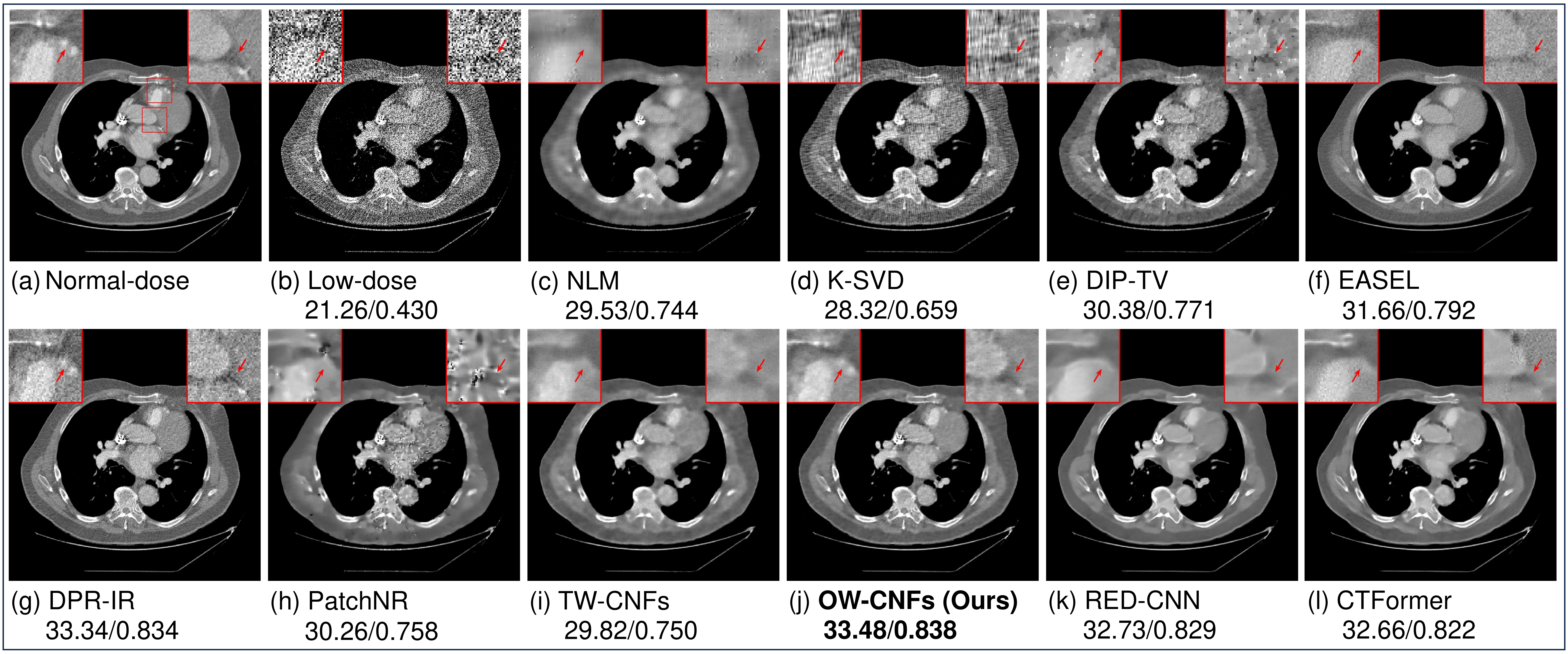}
    \vspace{-0.6cm}
	\caption{Reconstructed image results of each method on the \textit{"LIDC-IDRI"} dataset at dose $I_{0}=1\times{10^4}$. The display window is $[200,1430]$HU.}
	\label{fig_6_CA_1}
\end{figure*}

We chose two traditional image denoising methods NLM \cite{NLM_buades2011non} and K-SVD \cite{KSVD_elad2006image}, an unsupervised deep-learning method DIP-TV \cite{DIPTV-liu2019image}, an NFs-based regularization reconstruction method PatchNR \cite{PatchNR_altekruger2023patchnr}, two diffusion-model-based iterative reconstruction methods EASEL \cite{EASEL_he2022iterative} and DPR-IR \cite{DPR-IR_xia2023diffusion}, and two popular supervised LDCT image denoising methods RED-CNN \cite{REDCNN_chen2017low} and CTFormer \cite{wang2023ctformer}, as the comparative methods. We also ran the two-way iterative reconstruction algorithm \cite{dunfs_wei2022deep} introduced in Section 2.A with the same CNFs as in our method, here we use ``TW-CNFs'' to refer to it. As the evaluation indicators, we selected the commonly used PSNR and SSIM \cite{PSNR_hore2010image} for image quality assessment. Codes for all the comparative methods are publicly available. In the training phase of each method, we used the loss function and optimizer as described in the original training method, and selected the proper learning rate and mini-batch size. In the conditionalization part of our proposed OW-CNFs, we chose BM3D \cite{BM3D_dabov2007image} as the denoiser $\mathcal{D}$. All the experiments were performed on a server running Ubuntu 20.04.5 with Python 3.11, PyTorch 1.12.1, cuda 11.3, with a Nvidia Tesla V100 GPU card.

\subsection{Results}
On both the RRM and LIDC-IDRI datasets, the proposed OW-CNFs shows outstanding performance. Figure \ref{fig_5_RRM} shows one group of the reconstruction results of each method on the RRM dataset at dose $I_{0}=1\times{10^3}$. The reconstruction result of the proposed OW-CNFs presents comparable quality to the state-of-the-art generative-model-based methods and even the supervised methods. As shown in the zoomed-in areas, traditional methods NLM and K-SVD do not perform well. The unsupervised method DIP-TV also does not work well due to its weak priors. The two diffusion-model-based methods EASEL and DPR-IR show good results in effectively reconstructing the stripes, but both present noise residues and EASEL is more serious. The NFs-based method PatchNR performs denoising well but causes many block artifacts in the reconstructed image. This may be because PatchNR learns priors in image patches rather than the whole image, thus the global information would be lost to some extent and it is prone to mistakenly process large noise blocks as image structures. TW-CNFs shows noise retention and structure deformation in the reconstructed image: in the zoomed-in area, the length of the lower stripe is incorrect and its right end is severely blurred and even missing. Although the two supervised methods RED-CNN and CTFormer demonstrate satisfactory denoising, it is worth noting that the proposed unsupervised OW-CNFs has comparable performance with these two supervised methods and is even superior in evaluation indicators. While achieving effective denoising, the proposed OW-CNFs accurately reconstruct every stripe in the image.

\begin{table}[t]
	\centering
	\caption{The average indicators (psnr(db)/ssim) of each method on the test sets of all the datasets.}
	\begin{tabular}{lcccc} 
		\hline 
        &\multicolumn{2}{c}{\textit{RRM}} &\multicolumn{2}{c}{\textit{LIDC-IDRI}} \\
		\textbf{Method} &\textbf{PSNR} &\textbf{SSIM} &\textbf{PSNR} &\textbf{SSIM} \\ 
        \cline{1-5} \\
		\textbf{SART} &20.22 &0.559 &20.31 &0.400 \\
        \textbf{NLM} &25.88 &0.872 &29.01 &0.714 \\
		\textbf{K-SVD} &26.23 &0.749 &27.76 &0.628 \\
        \textbf{DIP-TV} &24.93 &0.793 &29.36 &0.726 \\
		\textbf{EASEL} &32.81 &0.878 &29.75 &0.717 \\
        \textbf{DPR-IR} &37.79 &0.965 &32.80 &0.804 \\
        \textbf{Patch-NR} &29.19 &0.926 &29.26 &0.727 \\
        \textbf{TW-CNFs} &31.97 &0.928 &29.48 &0.730 \\
        \textbf{OW-CNFs} &\textbf{38.60} &\textbf{0.972} &\textbf{33.14} &\textbf{0.819} \\
        \textbf{RED-CNN} &37.30 &0.967 &32.52 &0.809 \\
        \textbf{CTFormer} &31.02 &0.912 &32.25 &0.795 \\
		\hline
	\end{tabular}
	\label{table_1}    
\end{table}

\begin{figure*}[!t]
	\centering
	\includegraphics[width=7.2in]{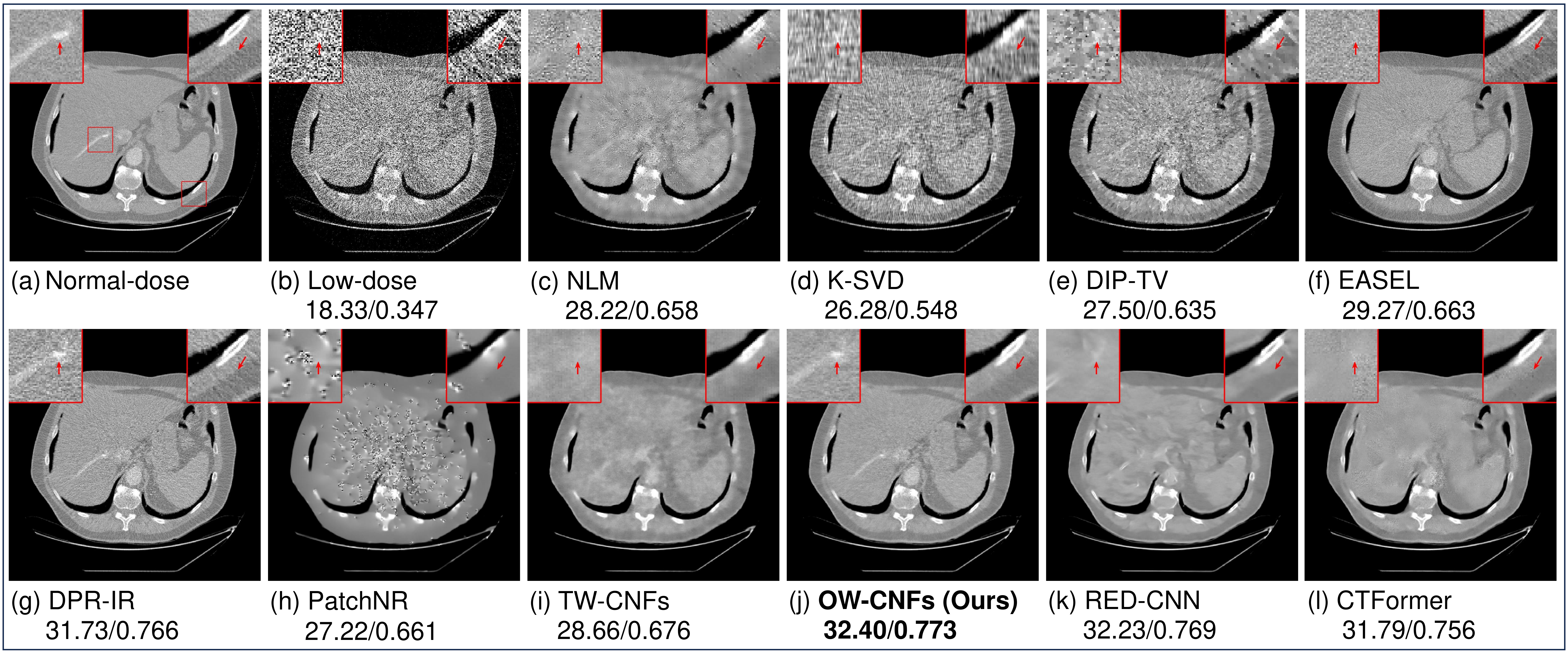}
    \vspace{-0.6cm}
	\caption{Reconstructed image results of each method on the \textit{"LIDC-IDRI"} dataset at dose $I_{0}=1\times{10^4}$. The display window is $[-100,1430]$HU.}
	\label{fig_8_CA_2}
\end{figure*}

\begin{figure*}[!t]
	\centering
	\includegraphics[width=7.2in]{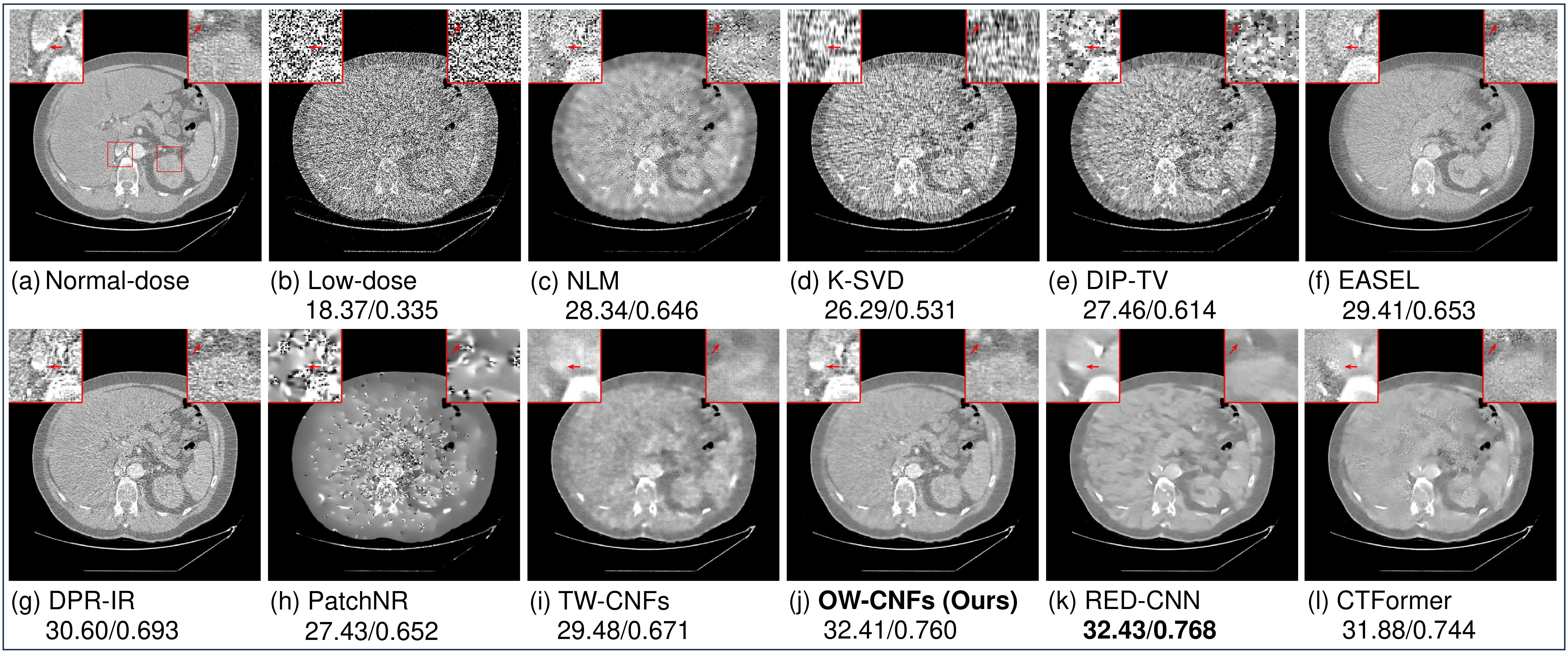}
    \vspace{-0.6cm}
	\caption{Reconstructed image results of each method on the \textit{"LIDC-IDRI"} dataset at dose $I_{0}=1\times{10^4}$. The display window is $[200,1130]$HU.}
	\label{fig_9_CA_3}
\end{figure*}

Figure \ref{fig_6_CA_1}, \ref{fig_8_CA_2} and \ref{fig_9_CA_3} show some comparisons of the reconstruction results on the LIDC-IDRI dataset at dose $I_{0}=1\times{10^4}$. On this dataset, the proposed OW-CNFs also presents outstanding performance. As shown in the zoomed-in areas, OW-CNFs demonstrates good accuracy on recovering image structures while achieving effective denoising. In contrast, although the diffusion-model-based methods EASEL and DPR-IR show good ability in recovering structures, they suffer from noise and artifacts, which might obscure image details. The NFs-based method PatchNR still shows a lot of pseudo structures in its results. The TW-CNFs shows more obvious structure deformation and severe loss of image details. Results of the two supervised methods RED-CNN and CTFormer show better contrast and line smoothing, however, they still cannot avoid distortion of structures and the detail loss in the reconstructed images. The average indicators on all the test sets are shown in Table \ref{table_1}, the proposed OW-CNFs has both the highest average PSNR and SSIM, even surpassing the two supervised learning methods RED-CNN and CTFormer. 

To test the reconstruction speed of the proposed OW-CNFs, we compared the average iteration number and reconstruction time of each generative-model-based method on the test sets. The images in the RRM dataset have a small size of $128\times128$, which allows us to directly train unconditional NFs and use the same one-way method for reconstruction. We also included this in the comparison and referred to it as OW-NFs. As shown in Table \ref{table_2}, our OW-CNFs has an excellent balance between performance and speed. As diffusion-model-based methods, EASEL and DPR-IR have to execute iterations on every noise level, which leads to more iterations and a longer reconstruction time. PatchNR has a fast speed because of its simpler way of regularization, however its performance is unsatisfactory. Although OW-NFs without the use of condition shows better indicators than OW-CNFs, iterating in the complete latent space makes it requires much more iterations and reconstruction time than OW-CNFs. In contrast, although the performance of the proposed OW-CNFs is slightly lower than the unconditional OW-NFs, it achieves good performance in much fewer iterations and less time, showing significant advantages in practicability. 

\begin{table}[t]
	\centering
	\caption{The average indicators (psnr(db)/ssim) and average inference time of each generative-model-based method on the test set.}
	\begin{tabular}{lccrr} 
		\hline 
		\textbf{Method} &\textbf{PSNR} &\textbf{SSIM} &\textbf{Iters} &\textbf{Time} \\ 
        \cline{1-5} \\
		\textbf{EASEL} &32.81 &0.878 &1800 &700s \\
        \textbf{DPR-IR} &37.79 &0.965 &1000 &405s \\
        \textbf{Patch-NR} &29.19 &0.926 &1102 &102s \\
        \textbf{OW-NFs} &\textbf{39.08} &\textbf{0.974} &1049 &1422s \\
        \textbf{OW-CNFs} &38.60 &0.972 &\textbf{55} &\textbf{49s} \\
		\hline
	\end{tabular}
	\label{table_2}    
\end{table}

\begin{figure}[!t]
	\centering
	\includegraphics[width=7.0cm]{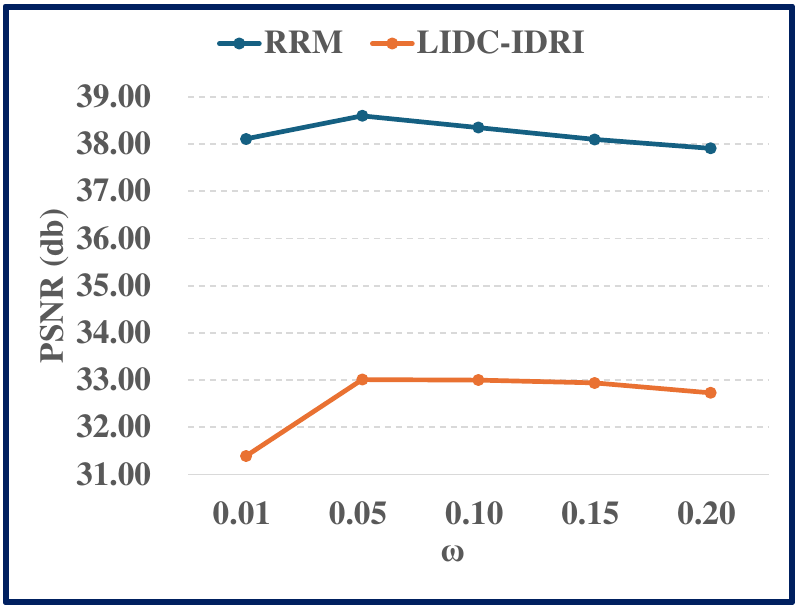}
    \vspace{-0.2cm}
	\caption{The trend of average PSNR under different $\omega$.}
	\label{fig_7_w}
\end{figure}

In our OW-CNFs, the relaxation parameter $\omega$ of OS-SART plays an important role in balancing the data fidelity and CNFs priors, and the change of $\omega$ will affect the reconstruction quality. We present the trend of average PSNR on the test sets under different $\omega$ in line charts, as shown in Figure \ref{fig_7_w}. Overall, the value of $\omega$ should be held in a small state, but a too small one will lead to bad data fidelity and reduce the accuracy of the reconstructed image. In the reconstruction process of the proposed OW-CNFs, we utilized the automatic backpropagation of \textbf{Pytorch} to adjust the $\omega$, by initializing it with a small value (e.g. $\omega=0.1$) and making the to-be-minimized equation (\ref{eq_3.17}) as the loss function. In this way, our reconstruction process can learn appropriate relaxation factors for different reconstructed images.

\section{Discussion and Conclusion}
We propose a novel unsupervised LDCT iterative reconstruction algorithm, OW-CNFs, based on conditional normalizing flows. By proposing an iterative reconstruction algorithm that conducts strict one-way transformation when conducting alternating optimization in the data and latent space, we solve the problems of detail loss and secondary artifacts that come from the current two-way transformation strategy. To perform unsupervised LDCT reconstruction with CNFs, we propose a conditionalization method for LDCT, thus achieving efficient training and fast reconstruction on high-resolution CT images. Moreover, we also proposed an efficient iterative reconstruction procedure by employing the incremental proximal minimization method and the linearization technique. With the incremental reconstruction of OS-SART, we realize simple and efficient computation of the iterations and effectively control the balance between the data fidelity and CNFs priors. Experimental results demonstrate our method's outstanding performance and fast reconstruction speed, in which our method achieves a high-level reconstruction speed among generative-model-based methods, and shows promising performance comparable to supervised learning methods.

As a generative-model-based method, the proposed OW-CNFs effectively avoids the main drawback of CNN-based denoising networks that easily cause over-smoothing and detail loss in CT images, and can produce images that more comply with diagnostic needs. Compared with diffusion-model-based methods, which also come from generative models, the main advantage of our OW-CNFs is that we do not need to strictly implement a fixed number of iteration steps that may be up to one thousand thus significantly reducing the reconstruction time, this should be attributed to our conditionalization method, by which the additional priors provided by the condition simplify the iterative process greatly. In fact, we are not the first to use CNFs in LDCT problems, but we are the first to apply CNFs effectively and deeply to the unsupervised LDCT reconstruction procedure. Compared to other methods that only use CNFs in image-domain post-processing and require lots of paired images for training, our method realizes the integration of the priors learned by CNFs into the reconstruction procedure of LDCT, avoiding the image-data inconsistency problem easily caused by post-processing networks. In addition, our method is an unsupervised framework that only requires normal-dose images for training, presenting higher application values in practical CT fields where training data is lacking, especially paired data.

Despite the promising performance, the proposed OW-CNFs still could be further improved. First, OW-CNFs introduces multiple hyperparameters and thus tuning work is a burden. Automatic adjustment would be a good way to solve this issue, like already done with the relaxation parameter $\omega$. But how to process other hyperparameters needs further research. Second, the training and generation process of the CNFs highly depends on the quality of the given conditions. How to generate conditions with higher accuracy and stronger priors for LDCT reconstruction tasks remains a problem worth studying. Although we propose a simple and well-working conditionalization method, it still relies on simple quantitative indicators and manual adjustment. Using more advanced information extraction methods such as pre-trained feature learning networks for conditionalization might improve the performance and speed of our method.

\IEEEpeerreviewmaketitle

\section*{Acknowledgment}
This work was supported by Beijing Natural Science Foundation (No.Z210003), National Natural Science Foundation of China (NSFC) (61971292) and China Scholarship Council (CSC) (No.202307300001). The authors are also grateful to Beijing Higher Institution Engineering Research Center of Testing and Imaging for funding this research work.

\bibliography{sec_bib_OW-CNFs}
\bibliographystyle{IEEEtran}

\end{document}